\documentclass{edm_article}
\usepackage{booktabs}
\renewcommand{\email}[1]{{\normalsize #1}}
\begin{document}

\title{A Methodological Framework for Capturing Cognitive-Affective States in Collaborative Learning}
% Submissions for EDM are double-blind: please do not include any author names or affiliations in the submission. 
% Anonymous authors:
\numberofauthors{3}

\author{
  \alignauthor
  Sifatul Anindho\\
  \affaddr{Colorado State University}\\
  {\small \email{sifatul.anindho@colostate.edu}}
  % \email{sifatul.anindho@colostate.edu}
  \and
  \alignauthor
  Videep Venkatesha\\
  \affaddr{Colorado State University}\\
  {\small \email{videep.venkatesha@colostate.edu}}
  \and
  \alignauthor
  Nathaniel Blanchard\\
  \affaddr{Colorado State University}\\
  \email{nathaniel.blanchard@colostate.edu}
}

\maketitle

\begin{abstract}
Identification of affective and attentional states of individuals within groups is difficult to obtain without disrupting the natural flow of collaboration. Recent work from our group used a retrospect cued recall paradigm where participants spoke about their cognitive-affective states while they viewed videos of their groups. We then collected additional participants where their reports were constrained to a subset of pre-identified cognitive-affective states. In this latter case, participants either self reported or reported in response to probes. Here, we present an initial analysis of the frequency and temporal distribution of participant reports, and how the distributions of labels changed across the two collections. Our approach has implications for the educational data mining community in tracking cognitive-affective states in collaborative learning more effectively and in developing improved adaptive learning systems that can detect and respond to cognitive-affective states.
\end{abstract}

\keywords{
    Collaborative Learning, Cognitive-Affective States, Affective Computing,
    Affect-aware Learning,
    Adaptive Learning Systems 
    
\section{Introduction}
Recently, researchers in educational data mining have been interested in modeling group dynamics in collaborative learning environments to understand how students interact, learn, and solve problems together \cite{sun2020towards,bradford2023automatic, khebour2024common, vanderhoeven2024multimodal, vrzakova2020focused, chandler2024computational}.
Here, we present initial experiments with retrospective cued recall paradigm for identifying cognitive-affective states of individuals in collaborative groups. The learning science community has recognized the importance of these states, highlighting their connection to attention, memory, problem solving, as well as motivation \cite{pekrun2002academic}. However, capturing these states in collaborative learning is difficult for a multitude of reasons, including 1) in-the-moment identification of states would disrupt the natural flow of collaboration, 2) group interaction creates complex emotional states that change rapidly due to various factors \cite{aoyama2022being}, and 3) students may hide their true emotions via social masking \cite{d2015review,underwood1997peer}. Our approach combines video cues and self-report surveys to reduce both recall bias and annotation effort, and the paradigm captures temporal details such as onset and duration, providing details of cognitive-affective state dynamics and their relationship to specific events or tasks. 

\section{Related Work}

Traditional methods of tracking cognitive-affective states in learning include self-reports \cite{pekrun2002academic, pekrun2014self, pekrun2016using, d2017advanced}, think-aloud protocols \cite{cotton2006reflecting}, and the use of behavioral observations by human coders \cite{bosch2016using, d2012towards}. Self-reporting is where individuals provide information about their emotional experiences through surveys, questionnaires, or interviews \cite{paulhus2007self}. Self-report measures during the task can disrupt the natural learning flow and are restricted to evaluating conscious thoughts, while similar measures after the task will be subjected to recall bias and lack temporal resolution \cite{pekrun2020self}. Think-aloud measures, in which learners verbalize their feelings during learning \cite{eccles2017think}, can also interfere with natural emotional processing and be challenging to implement in collaborative settings \cite{cotton2006reflecting}. 
% Physiological and behavioral measures such as electrodermal activity, heart rate, skin temperature, and facial action units can objectively detect emotional states \cite{bosch2016using,bixler2015automatic,monkaresi2016automated, febriantoro2023promise}. However, external factors such as environmental conditions and physical activity can impact their reliability \cite{febriantoro2023promise}. These sensors also face challenges in real-world educational settings due to high costs, technical complexity, and privacy concerns \cite{d2012towards}. 
Behavioral observation and coding by human judges involves trained coders annotating live interactions or recordings for specific cues \cite{hailpern2009a3}, but it is labor intensive, requires ensuring coder consistency, and behaviors can sometimes be difficult to map to cognitive-affective states \cite{Heyman_Lorber_Eddy_West_2014}. Achieving high agreement among coders is also difficult, and the process can be slow and prone to coder fatigue \cite{hayes2022identifying}.

In 2009, D'Mello and Graesser \cite{d2011half} used a combination of self-report and behavioral observation in which learners watched and judged their own session with an intelligent tutor immediately after taking a test.
% More recently, Anindho et al. implemented a similar method where students individually watch a video of their collaborative task and report their thoughts at specific moments. 
This approach provides a low-cost, non-intrusive method that minimizes some of the limitations of self-reporting, think-aloud and behavioral observations alone. 
% We use a similar framework that incorporates self-reporting with retrospective video-stimulated recall in collaborative learning.
% Our research question is: Can combining self-caught and probe-caught methods with retrospective video-stimulated recall effectively capture cognitive-affective states in collaborative learning?
% \section{Related Work}
More recently, Anindho et al. \cite{anindho2025exploration} used a similar framework to explore the vernacular of internal monologues in a collaborative task by asking participants to review session recordings and self-report their internal feelings. Four-person groups completed a Lego task, then watched a video of it while sharing their thoughts and feelings about their performance. The monologues were recorded through screen recording which also captured the corresponding video segments. The analysis revealed distinct patterns in n-grams and manually extracted keywords relevant to the learner's internal states. They thematically identified common cognitive-affective labels in collaborative problem-solving using those keywords and followed up with a semantic similarity evaluation using cosine similarity. Figure \ref{histogram_label_occurences} shows the frequency of words associated with each emotion theme in this study.
% \subsection{Proposed Framework}

This study extends Anindho et al.'s retrospect cued recall paradigm \cite{anindho2025exploration} by combining video playback with self-report surveys to capture a subset of these states (with the addition of `Curious') in a structured way. Participants watch a video recording of their task session in an interactive application and are prompted to report their cognitive-affective states at specific points, either through self-caught reporting (participant chooses to open the survey) or probe-caught reporting (the survey automatically opens after a set time interval of idleness). When the survey opens, participants select options that best describe their cognitive-affective states and indicate whether a particular state is onset or ongoing, with the option to estimate the duration of ongoing states. Due to the overlapping nature of cognitive-affective states \cite{corradi2014cognitive}, participants can report multiple states at the same time without any restrictions. 

\begin{figure}
\centering
\includegraphics[width=0.5\textwidth]{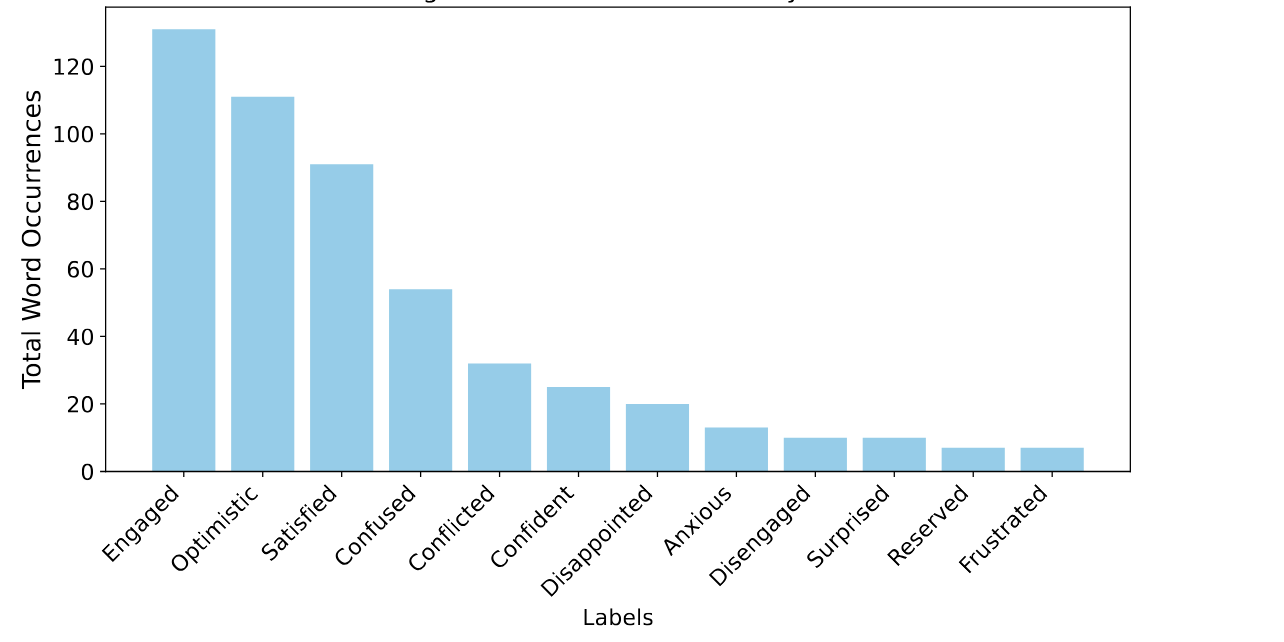}
\caption{The total occurrences of words associated with each emotion theme label from Anindho et al. \cite{anindho2025exploration}} \label{histogram_label_occurences}
\Description[The total occurrences of words associated with each emotion label. In descending order: Engaged, Optimistic, Satisfied, Confused, Conflicted, Condifent, Dissapointed, Anxious, Disengaged, Surprised, Reserved, Frustrated]{}
\end{figure}

\section{Methods}
\subsection{Collaborative Task Experiments}
\subsubsection{Participants}
We recruited a total of 27 participants, organized into 9 groups of 3 individuals each. Participants were required to be at least 18 years
old and able to speak in English. Recruitment efforts were focused in the department where the authors are appointed, and potential participants were informed of the study through verbal communication. Participants were compensated for their time with 15 US Dollars. Table \ref{tab:dem_data} presents the demographics of our participants. 
% Our participants were mostly male (18) and female (7), with 2 identifying as non-binary. The age range was primarily 18-24 (14) and 25-34 (11), with 2 participants aged 35 years or older. Most of the participants spoke English (17) as their native language, while other native languages represented included Tamil (2), Bengali (2), and 5 other languages. The ethnic breakdown was 13 Asian, 12 White, 1 Hispanic, and 1 participant of two or more ethnicities.

\begin{table}[ht]
\centering
\caption{\textbf{Demographics of participants from the collaborative task experiments}}\label{tab:dem_data}
\begin{tabular}{lccc}
\toprule
\textbf{Gender} & Male & Female & Non-binary \\
 & 18 & 7 & 2 \\
\midrule
\textbf{Native Language} & English & Other & \\
 & 17 & 10 & \\
\midrule
\textbf{Age} & 18--24 & 25--34 & 35+ \\
 & 14 & 11 & 2 \\
\midrule
\textbf{Ethnicity} & Asian & White & Hispanic / Two+ \\
 & 13 & 12 & 2 \\
\bottomrule
\end{tabular}
\end{table}
\subsubsection{The Collaborative Task}
Participants completed the first part of the Weights Task, a collaborative problem-solving activity introduced by Khebour et al. [20], where they used a balance scale to determine the weights of five blocks with one of known weight, submitting one group response via survey. Each group completes one survey. In this variation of the task, an AI agent was also present in some groups to introduce constructive friction that makes the team pause and reflect when appropriate.

\subsubsection{Experiment Setup}
The study setup included an Azure Kinect camera, three microphone headsets, three Lenovo tablets, a laptop with a webcam, and three SkullCandy HeshEvo headphones. The physical environment consisted of a round table with three chairs, a sound-enabled display, and a Weights Task set. This task set included five blocks of different weights (ranging from 10g to 50g), a balance scale and a worksheet to keep track of answers.
\begin{figure}
    \centering
    \includegraphics[width=1\linewidth]{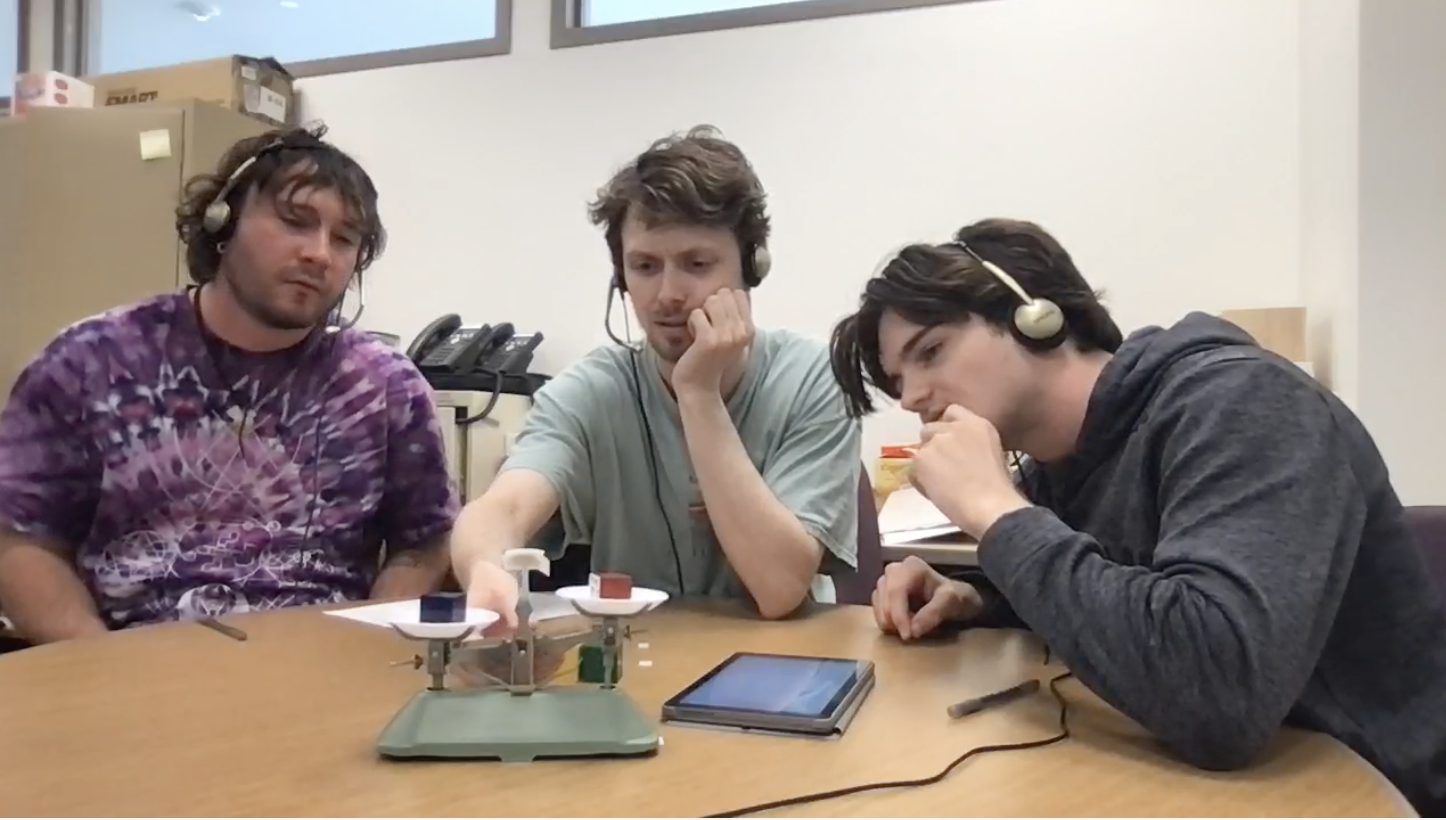}
    \caption{The physical environment of the experiment (The sound enabled display is behind the POV of the scene)}
    \label{fig:physical-setup}
    \Description[The physical environment of the experiment (The sound enabled display is behind the POV of the scene)]{Image contains 3 participants collaborating on the Weights Task, the Weights Task Set, a tablet}
\end{figure}
Figure \ref{fig:physical-setup} shows an image of the environment.

% These experiments were conducted in conjunction with a different study examining the impact of an AI assistant on small group work. Groups were assigned to either an AI-assisted condition or a non-AI condition. The AI assistant's objective was to detect and mitigate friction in collaborative dialogue by providing timely and relevant interventions.
\subsubsection{Experimental Procedure}
Before the experiment began, participants provided their consent and video recordings were
initiated. Participants were provided instructions on the task and asked to begin.
Once the task was finished, they completed a demographic information survey on one of the tablets.
Afterwards, they completed the self-report survey using our extended framework outlined in Section 2. The set time interval for the probe-caught survey was 60s. 
The label options on the survey were “Confused”, “Disengaged”, “Curious”, “Optimistic”, “Frustrated”, “Conflicted” and “Surprised”. If the provided options did not accurately capture their experience, participants could choose ``Other” and enter a description of their cognitive-affective state. The ``Other" option was added after the second experiment to capture any commonly mentioned labels not among the initial 7, so groups 1 and 2 did not have this option. For each report, metadata including timestamp, video time, labels, report type, group ID, participant ID, and participant role were collected.

Cognitive-affective state reports collected from the participants were classified into two collection methods: self-caught and probe-caught. We did this by finding the time differences between consecutive reports. After each self-report, the video skipped forward by 1 second before resuming to ensure that participants were not immediately re-probed after a survey. In the probe-caught method, the system automatically prompted participants to report their cognitive-affective state if 60 seconds had passed since the start of the video review without a self-report, or every 61 seconds after a previous report. Therefore, any report with a time difference that aligns exactly with the probe-frequency is classified as probe-caught; the rest are self-caught.

This study was conducted in accordance with the Institutional Review Board (IRB) approval and guidelines. Prior to participation, all participants provided informed consent, ensuring they understood the study's procedures, risks, and benefits. Video recordings collected during the study are stored on a secure server at our institution, where they will be retained for a period of 5 years in accordance with our data management policies.

\subsection{Data Preprocessing and Analysis}
Data preprocessing consisted of standardizing the case of labels across all participants, expansion of multi-label responses into separate entries, and removal of reports with no labels. For groups 3-9, ``Other" responses were processed by capitalizing single-word responses and adding them to the label list. Timestamps were normalized within each group by scaling them relative to the length of the video for that group. 

% Time windows for each reported cognitive-affective state were estimated based on onset and duration information. For onset states, the window spanned from 10 seconds before the report time to 10 seconds after the report time. For ongoing states, the window started 10 seconds before the state onset and ended 10 seconds after the report time. Overlapping windows of the same label were merged for each participant. The calculations were adjusted to ensure nonnegative start times.
The frequency of each reported cognitive affective state was calculated between all participants. The most frequently reported states were identified and frequency distributions were created for all labels. The frequency distributions between self-caught and probe-caught reports were compared and visualized, and descriptive statistics (mean, standard deviation, and standard error) were found for each type of report.
% We found the average time, mean and standard deviation to observe how often participants reported their cognitive-affective state using both methods.
% Comparative visualizations were created to illustrate differences in frequency distributions between self-caught and probe-caught reports.
% We also conducted a group-level analysis by calculating the frequency of cognitive-affective states reported by each group and identifying the top 5 most frequent states. We also measured task performance using objective metrics, including total task time and subjective performance ratings collected through post-task surveys, where participants rated their performance on a 7-point Likert scale ranging from Extremely Poor to Excellent. 
% These metrics allowed us to compare group performance and identify potential patterns between cognitive-affective states and task outcomes.
The distribution of reports was then analyzed across the normalized task time. The timing of specific state reports during the task was tracked, and scatterplot visualizations were created to illustrate label occurrences over normalized time.\footnote{Source code: https://anonymous.4open.science/r/edm-2025-mmla-submission-7776/README.md}.
\section{Results}

\begin{table}[ht]
\centering
\caption{\textbf{Statistics on the number of Labels of Cognitive-Affective States Reported}}\label{tab:report-stats}
\begin{tabular}{lccc}
\toprule
 & All Labels & Self-Caught & Probe-Caught \\
\midrule
Total Count & 359 & 129 & 230 \\
Mean & 13.30 & 4.78 & 8.52 \\
SD & 4.67 & 4.81 & 5.69 \\
\bottomrule
\end{tabular}
\end{table}

Table \ref{tab:report-stats} shows the descriptive statistics on the number of labels of cognitive-affective states reported. 
% A total of 359 label reports were collected, with 129 being self-caught and 230 probe-caught. On average, participants reported 13.30 labels, with self-caught labels averaging 4.78 and probe-caught labels averaging 8.52. The standard deviation for all labels was 4.67.
\begin{figure}
    \centering
    \includegraphics[width=0.9\linewidth]{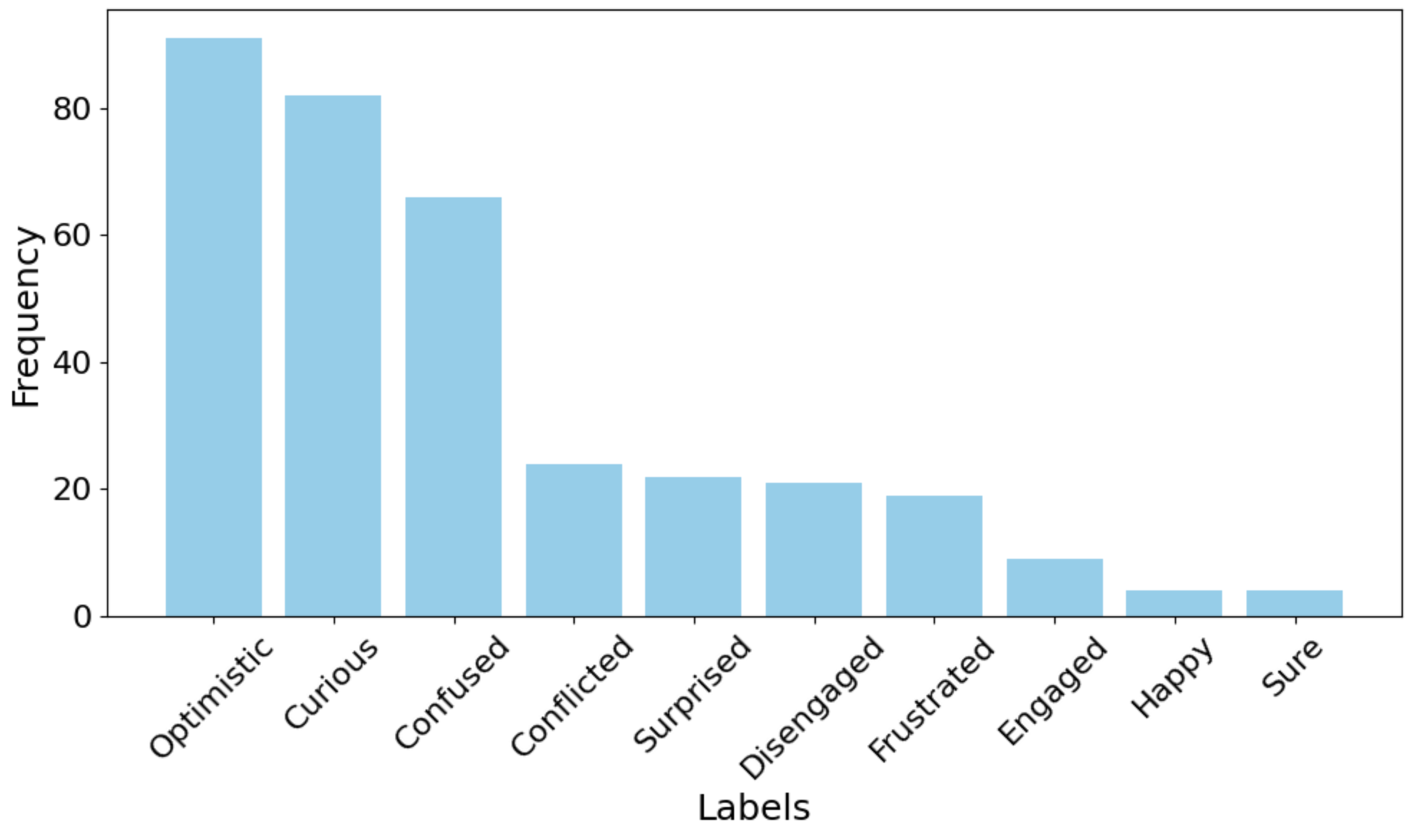}
    \caption{Distribution of the 10 most frequent labels reported in this study using both self-caught and probe-caught}
    \label{fig:overall-label-frequencies}
    \Description[Distribution of the 10 Most Frequent Labels Collected Overall]{A bar chart showing the overall distribution of the 10 most frequent labels reported in this study using both methods}
\end{figure}
Figure \ref{fig:overall-label-frequencies} shows the overall distribution of the 10 most frequent labels in our data. The most frequent labels were ``Optimistic" (91), followed by ``Curious" (82) and ``Confused" (66). Other notable labels included ``Conflicted" (24), ``Surprised" (22), and ``Disengaged" (21), while ``Frustrated" appeared 19 times. Less frequent labels such as ``Engaged" (9), ``Happy" (4), and ``Sure" (4) were part of the ``Other" category, where participants typed their own state labels.
\begin{figure}
    \centering
    \includegraphics[width=0.9\linewidth]{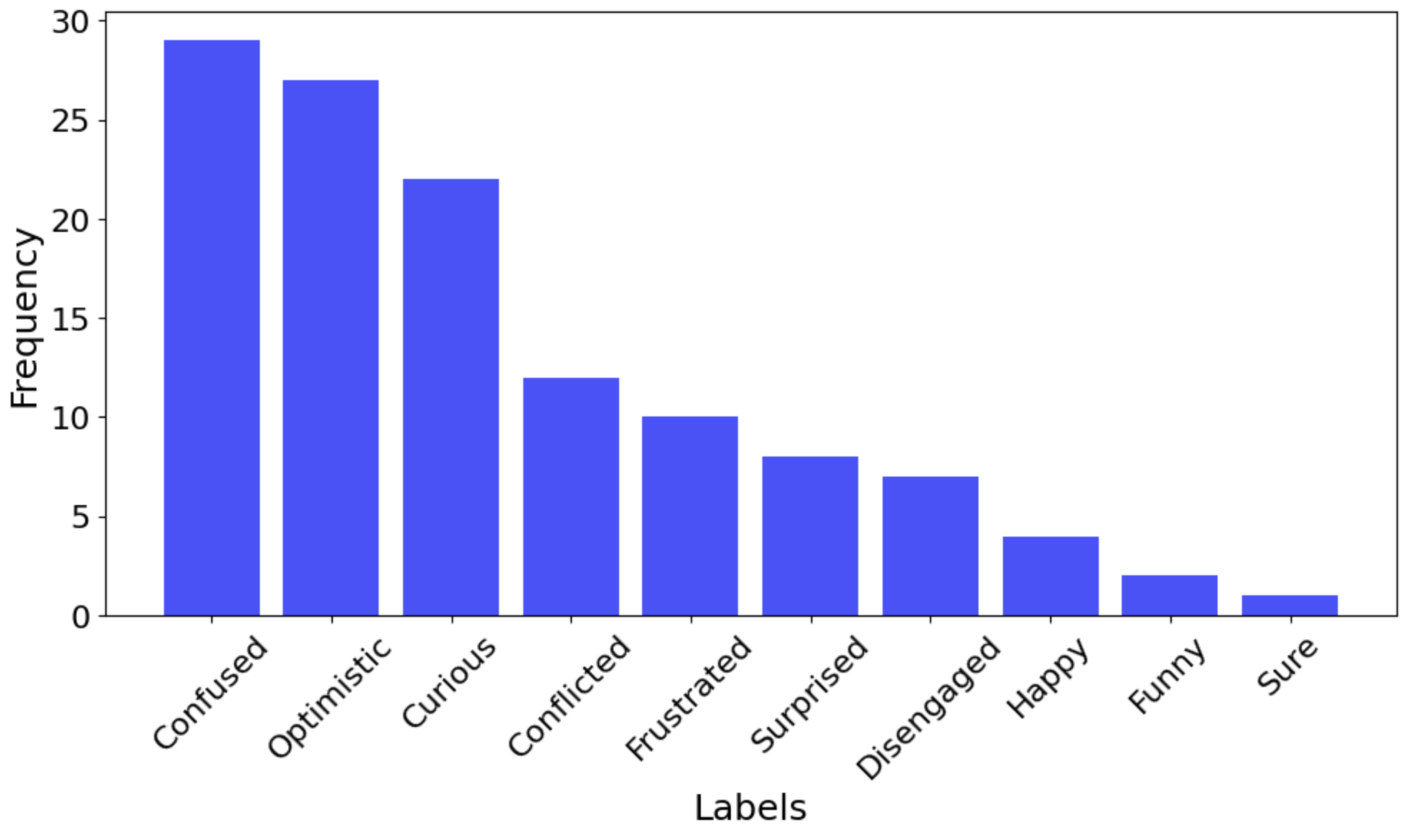}
    \caption{Distribution of the 10 most frequent labels collected using self-caught}
    \label{fig:self-caught-top-10}
    \Description[Distribution of the 10 Most Frequent Labels Collected Using Self-caught]{A bar chart showing the overall distribution of the 10 most frequent labels reported in this study using self-caught}
\end{figure}
\begin{figure}
    \centering
    \includegraphics[width=0.9\linewidth]{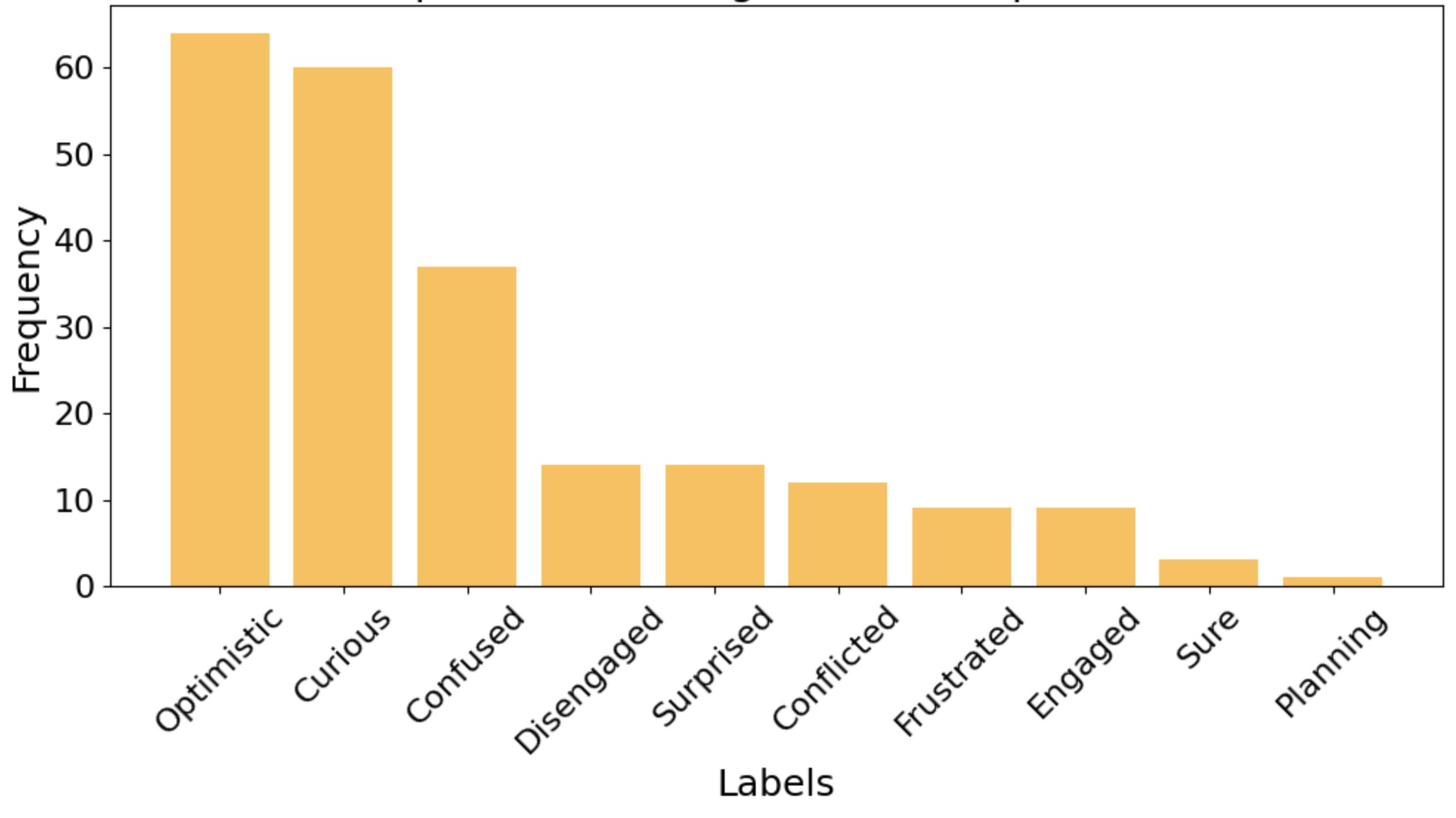}
    \caption{Distribution of the 10 most frequent labels collected using probe-caught}
    \label{fig:probe-caught-top-10}
    \Description[Distribution of the 10 Most Frequent Labels Collected Using Probe-caught]{A bar chart showing the overall distribution of the 10 most frequent labels reported in this study using probe-caught}
\end{figure}
\begin{figure*}
    \includegraphics[width=1\linewidth]{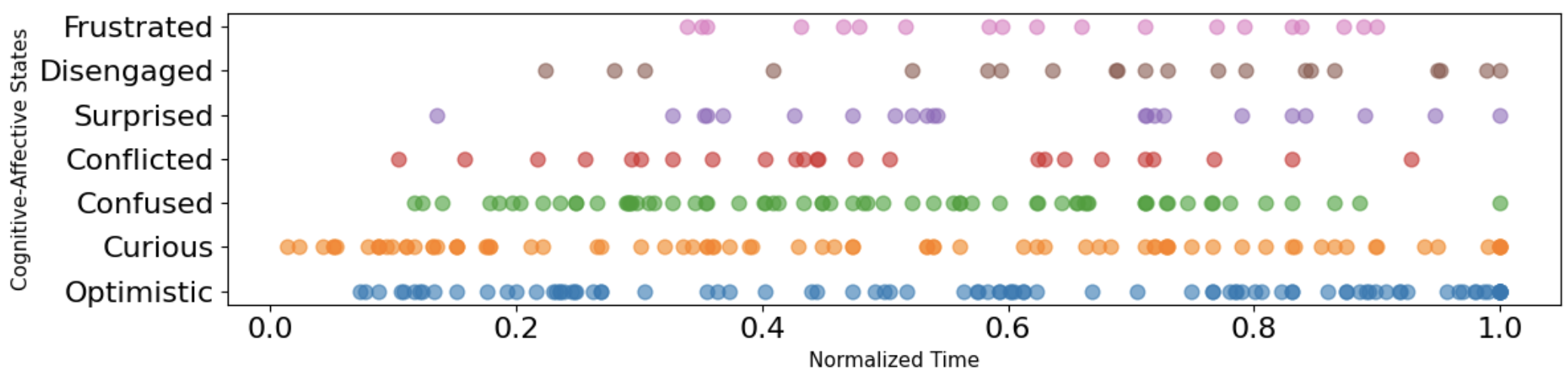}
    % \centering
    \centering
    \caption{A scatter plot showing the distribution of the 7 most frequent cognitive-affective state reports over normalized task time}
    \label{reports-normalized-time}
    \Description[7 Most Frequent Cognitive-Affective State Reports Over Normalized Task Time]{A scatter plot showing the distribution of the 7 most frequent cognitive-affective states reports over normalized task time}
\end{figure*}
\begin{table}[ht]
\centering
\caption{Descriptive statistics of the time difference between the reports}
\begin{tabular}{lcc}
\toprule
Metrics & Value \\
\midrule
Average Time Difference (seconds) & 40.43 \\
Standard Deviation (seconds) & 25.45 \\
Standard Error (seconds) & 1.34 \\
\bottomrule
\end{tabular}
\label{tab:time-difference}
\end{table}
Figures \ref{fig:self-caught-top-10} and \ref{fig:probe-caught-top-10} shows the distribution of the 10 most frequent labels collected using self-caught and probe-caught respectively. For self-caught labels, the most frequent were ``Confused" (29), ``Optimistic" (27), and ``Curious" (22). The probe-caught labels showed a similar top-three ranking, with ``Optimistic" (64), ``Curious" (60), and ``Confused" (37) being the most common. None of the labels typed in the ``Other" section were more frequent than the seven initial labels in self-caught or probe-caught reports. Table \ref{tab:time-difference} shows the descriptive statistics of the time difference between the reports.

Figure \ref{reports-normalized-time} shows the labels reported by all participants during the normalized task time. ``Curious" reports were common at the start, while ``Confused" reports started shortly after. ``Disengaged" reports increased in the latter half, and ``Frustrated" reports increased as the task progressed, but then dropped off shortly before completion. ``Surprised" labels were intermittent in nature throughout the task.
\section{Discussion}
 % Our initial results offer a proof-of-concept for the effectiveness of this method to gather information about cognitive-affective states in collaborative learning.

The distribution of states was largely consistent between self-caught and probe-caught reports, with minor deviations. The prevalence of ``Optimistic" and ``Curious" reports may suggest that participants were in a neutral-engaged state \cite{d2012dynamics}, and co-opted one of these labels as the closest approximate for that state. In the original collection by Anindho et al. \cite{anindho2025exploration}, participants often did not speak for some portions of the video --- in retrospect, this may correspond with the neutral-engaged state, indicating an additional label explicitly for this should be made available to participants. We also note that participants reported states such as ``Confused" with greater frequency when self-reporting, and reported more low-arousal emotions when probed, further indicating the use of these labels as stand-ins for a neutral-engaged state. 

The temporal analysis revealed that participants tend to be more disengaged as the task progressed, which can be explained by fatigue, waning interest or mind wandering \cite{zanesco2024mind}. ``Confused" reports persisted over longer periods without interruption, while ``Frustrated" reports peaked mid-task, aligning with the established theory that confusion is a sustained state \cite{d2014confusion} and frustration is an intermediate response \cite{d2012dynamics} that follows as challenges continue. ``Surprised"  was intermittent and transitory \cite{baker2010better}. 

\section{Limitations and Future Work}
This study's small dataset and broad category of labels in this study limited the scope of the analysis and the experiences of the participants \cite{riener2020addressing}. Additionally, the results may not generalize to other collaborative learning tasks, and validation of a different task may be necessary to confirm findings \cite{highhouse2009designing}. The temporal fidelity of self-reports also remains undetermined, since the method has not been compared to the behavioral and physiological measurements of our data. In many ways, this study amounts to a pilot methodology and identifies strengths, weaknesses, and directions to refine the methods for future collections. 
Future work will involve additional experiments with different collaborative learning tasks and fine-grained analyses to further validate the methodology. Label selection will be iteratively improved based on feedback and observation of trends. We will also analyze affective chronometry \cite{davidson1998affective} \cite{rosenberg1998levels} in collaborative learning by investigating the temporal patterns of cognitive-affective states and exploring methods to map these reports to specific task windows with behavioral and physiological markers. Future work will also investigate the likelihood of transitioning between states \cite{d2011half}.

\section{Conclusion}
This study investigated a retrospective cued recall paradigm to capture cognitive-affective states in collaborative learning, combining self-caught and probe-caught methods with retrospective video-stimulated recall. Our initial findings provide insights into the dynamics of these states and their distributions using different reporting methods, which are consistent with previous research, but current limitations highlight the need for further validation and refinements. Our approach can help the educational data mining community develop more efficient, reliable, and low-disruption methods for tracking cognitive-affective states and creating adaptive learning systems that respond to them, enabling educators to better support students' emotional and cognitive needs in collaborative learning. Future work will focus on replicating these findings in diverse contexts, incorporating multimodal behavioral and physiological indices, and developing affective chronometry models to understand the temporal dynamics of cognitive-affective states in collaborative learning. 
\bibliographystyle{abbrv}
\bibliography{sigproc}

\end{document}